\documentstyle[11pt,newpasp,twoside,psfig]{article}
\def\ea{et al.}
\def\rosat{{\sl ROSAT}}
\def\asca{{\sl ASCA}}
\def\chandra{{\sl Chandra}}
\def\xmm{{\sl XMM-Newton}}
\def\as{$^{\prime\prime}$}

\def\g21{G21.5$-$0.9}

\def\edcomment#1{\iffalse\marginpar{\raggedright\sl#1\/}\else\relax\fi}
\reversemarginpar

\begin{document}
\title{The Intriguing Plerionic Supernova Remnant: G21.5-0.9}
\author{Safi-Harb, S.\altaffilmark{1}}
\affil{U. of Manitoba, Department of Physics and Astronomy, 515 Allen Bldg., Winnipeg, MB, R3T 2N2 Canada}
\altaffiltext{1}{NSERC fellow; samar@physics.umanitoba.ca}
\author{Harrus, I. M., Petre, R.}
\affil{NASA/GSFC, Code 662, Greenbelt, MD 20771, USA}
\author{Pavlov, G. G., Koptsevitch, A. B., \& Sanwal, D.}
\affil{Penn. State University,  525 Davey Lab, PA, 16802, USA}

\begin{abstract}
\g21 is a center-brightened (or plerionic) supernova remnant
(SNR) whose properties hint at the presence of a pulsar --
yet no pulsations have been found at any wavelength.
Early observations with \chandra\ led to the
discovery of an extended component,
making the SNR at least twice as big as originally thought.
Our analysis indicates that this low-surface brightness
extended component is non-thermal
with a filamentary hard structure in the
northern quadrant. We perform a spatially resolved spectroscopy and
find no evidence of line emission using a 72 ksec exposure with
ACIS-S.  The 5\arcmin\ diameter remnant is well fitted with a power law
with a photon index steepening from 1.5 to 2.7 ($N_H$=2.2$\times$10$^{22}$~cm$^{-2}$).
Using a 76 ksec exposure with the HRC, we derive an upper limit
of 16\% on the pulsed fraction from a putative pulsar.
We also infer the parameters of the `hidden pulsar' in
\g21. This remnant remains  unique and intriguing since it
is, to date, the only candidate whose size is bigger
in X-rays than in the radio.
\end{abstract}

\section{Introduction}
\g21 has a flat radio spectral index,
and is highly polarized (Green~2000).
It is a 1.3\arcmin\ diameter plerion ($\sim$~1.9~pc at
a distance of 5.0~kpc) in the radio, infrared,
and in X-rays (prior to \chandra; Becker and Szymkowiak 1981).
 Early \chandra\ observations (Slane \ea\ 2000) 
revealed a low-surface brightness
extended component interpreted as the missing SNR shell.
\g21 shares similar properties to 3C 58 and CTB~87,
plerions having a spectral break at low frequencies
and classified as plerions of the {\it second kind}
(Woltjer \ea\ 1997).
Bock, Wright, \& Dickel (2001) have however recently questioned the
spectral break in \g21.
In this paper, we highlight the \chandra\ observations.
More details about the analysis (including \rosat\ and \asca\
observations) and interpretation of our results can be found in Safi-Harb
\ea\ (2001).

\section{Imaging}
\begin{figure}[tbh]
\centerline{\psfig{figure=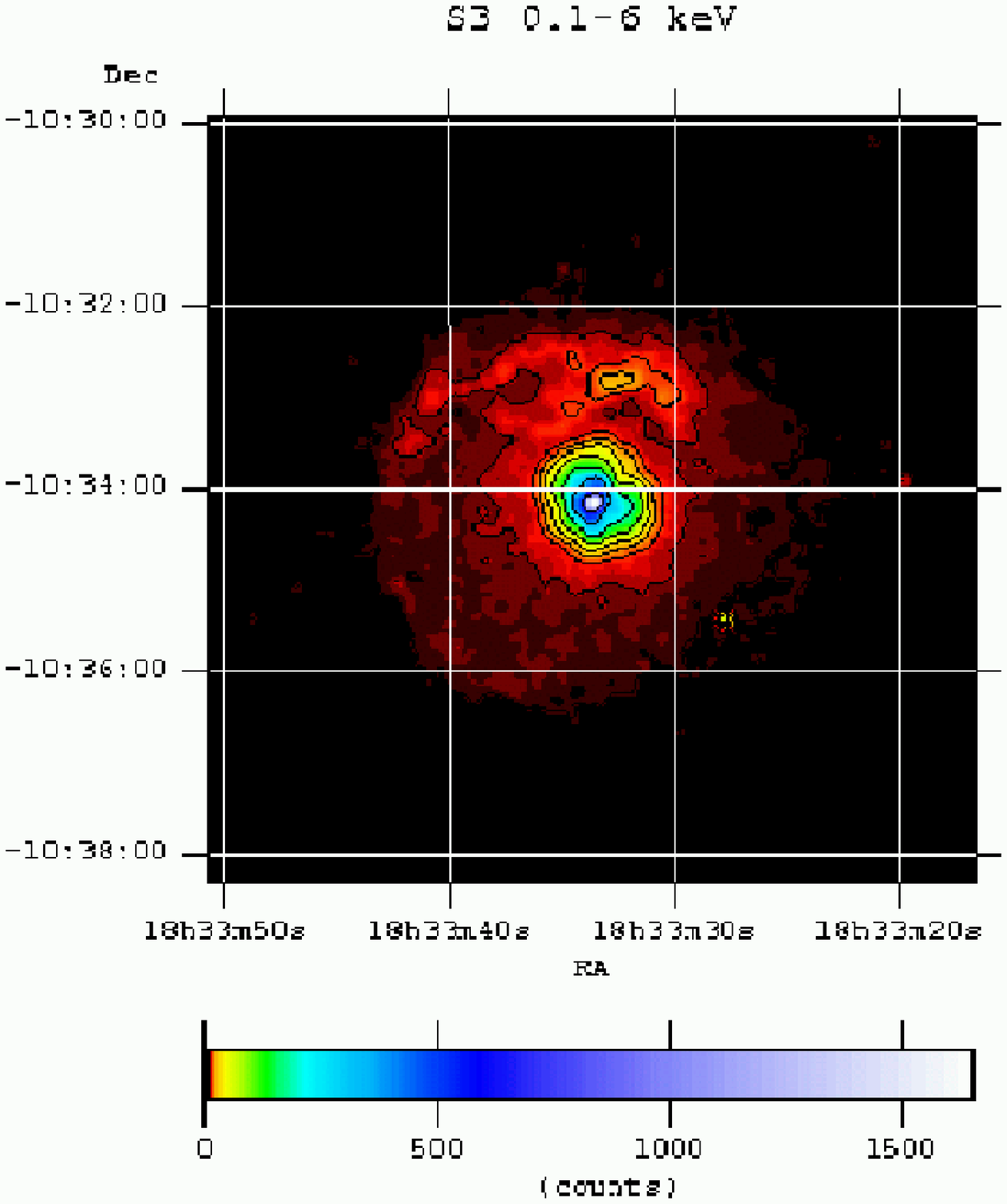,width=2.8in}
\psfig{figure=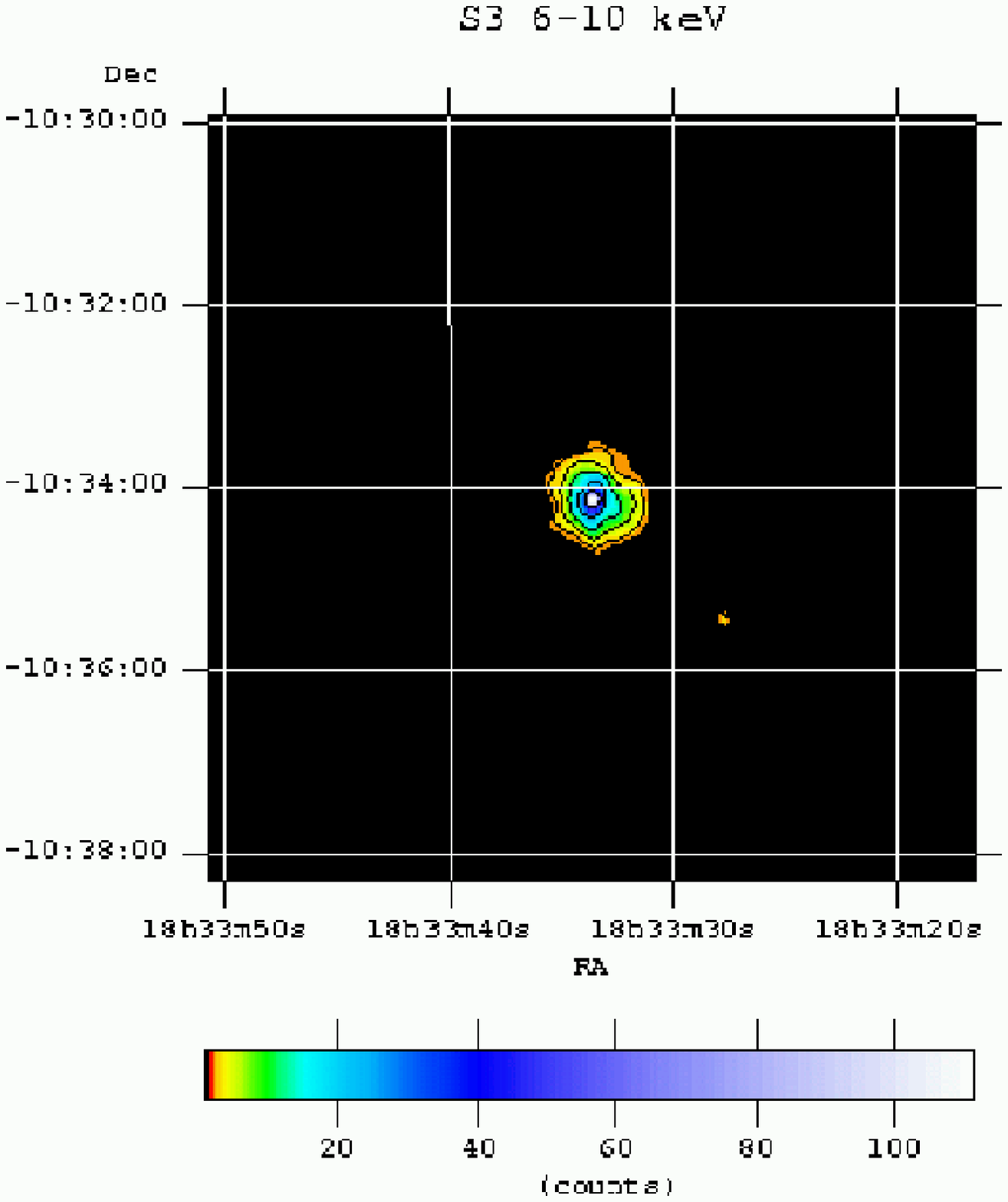,width=2.8in}}
\caption{Soft (Left) and
hard (Right) energy images of G21.5-0.9.
The extended component is evident in the soft band image.}
\end{figure}
In Fig.~1, we show the \chandra\ 
images of \g21 in the soft and hard energy bands.
 With {\it ROSAT} and previous radio observations,
only the bright 40\as \ radius core is detected.
\chandra\ allowed the discovery of
a faint extended component  (Fig.~1, left) making the SNR
at least twice as big as previously thought.
In Fig.~2, the hardness ratio map (2.4--10 keV over
0.5--2.4 keV) shows that the central core is harder than the 
extended component (in agreement with our spectral analysis below). Moreover, the extended component
is harder in the northern quadrant, with brighter knots and filamentary structures.

\begin{figure}[tbh]
\centerline{
\psfig{figure=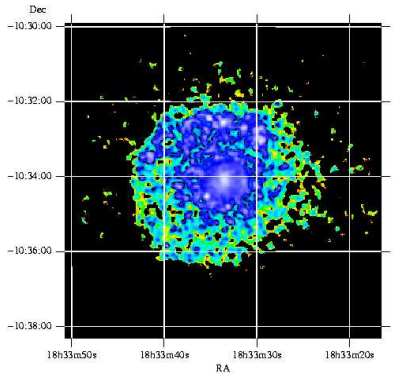,width=2.2in}}
\caption{ Hardness ratio map (2.4--10
over 0.5-2.4 keV) obtained with ACIS-S
and smoothed with a Gaussian with $\sigma$=3".
 The northern knotty quadrant is harder
than the rest of the extended component.}
\end{figure}

\section{Spatially Resolved Spectroscopy:}
We perform a spatially resolved
spectroscopy of the plerion using a 72 ksec exposure with
${\it Chandra}$ ACIS S3.
The spectrum of the inner core is extracted from a circle
of radius  $R$=40$^{\prime\prime}$ (with rings of 5" thickness).
For the  extended component, we
extract a spectrum from 50$^{\prime\prime}$--150$^{\prime\prime}$. 
All spectra are best described by a power law
with the photon index, $\Gamma$, steepening away from the bright center
from 1.5 to 2.7 ($N_H$=2.2$\times$10$^{22}$~cm$^{-2}$).
We rule out thermal models for the extended component.
Collisional equilibrium ionization (CEI) models, such as $\it{apec}$,
yield poor fits ($\chi_{\nu}^2$$\sim$1.15). Non-equilibrium ionization (NEI) models,
such as $\it{PHSOCK}$,
are unlikely since they
yield a very low ionization timescale, and an interstellar
column density, $N_H$, much lower than that derived for the
SNR (Table~1).
Furthermore, we find no evidence of line emission
from the filaments and knots. Their spectra are well 
fitted with a power law model.

\begin{table}[tbh]
\begin{center}
\caption{Spectral parameters of the fits to the low-surface-brightness
extended component using ACIS-S3 observations.}

 \begin{tabular}{c|ccc}
 Model &  $N_H$\tablenotemark{a} ($\times$10$^{22}$ cm$^{-2}$)
 & Model Parameter & 
 $\chi^2_{\nu}$ ($\nu$)\\
\tableline
  Power Law & 1.83 (1.77--1.90) & $\Gamma$ =  2.36 (2.30-2.43)  & 1.0 (658) \\
 ${\it PSHOCK}$ & 1.53 (1.48--1.57) & $kT$ = 3.50 (3.28--3.74) keV  & 0.86 (657) \\
            
 &                   & $\tau$\tablenotemark{b} = 1.1 (0.8--1.3)$\times$10$^9$ cm$^{-3}$~s  & \\
\end{tabular}
 \tablenotetext{a}{Best fit values are indicated. When freezing $N_H$ to 2.2~$\times$
10$^{22}$~cm$^{-2}$, the power law model gives $\Gamma$ = 2.73~$\pm$~0.04
($\chi^2_{\nu}$~=~0.98; $\nu$~=~659),
and thermal models are rejected ($\chi^2_{\nu}$$\sim$1.5)}
\tablenotetext{b}{The ionization timescale, $n_et$; where $n_e$ is
 the postshock electron density, and $t$ is the age of the shock}
\end{center}
\end{table}

\section{The putative pulsar}
The morphology and spectrum of the inner core indicate the
presence of a pulsar powering \g21. 
We did not find pulsations using the \asca\ and \chandra\ 
archival observations. Using 5 HRC observations totaling to 76 ksec,
we put an upper limit on the
pulsed fraction of 16\%. We also measure an unabsorbed
flux of 2.4$\times$10$^{-12}$~erg~cm$^{-2}$~s$^{-1}$
from the point source.
The failure to detect pulsations
could be due to a beaming effect.
Using various $L_X$--$\dot{E}$ empirical relationships
for pulsar-powered plerions, we find that a plausible 
estimate of the spin down energy loss, $\dot{E}$, is
$\dot{E}_{37}\equiv \dot{E}/(10^{37}~{\rm erg}~{\rm s}^{-1})\sim 3$--6.
In the Kennel and Coroniti (1984) model,
the pulsar wind gets shocked at a radius, $R_s$, given by equating the pressure of the
pulsar's wind with the pressure in the nebula.
Beyond this radius, a non-relativistic flow transports the plasma from
the shock region to the edge of the nebula.
The size of the nebula, $R_n$ is related to the shock radius,
$R_s$ as: $R_n$/$R_s$ $\sim$ 1/$\sqrt\sigma$; where
$\sigma$ is the so-called pulsar wind magnetization parameter defined as
the ratio of the Poynting flux to the particles flux .
Using the power law fit to the inner 40\as \ radius core,
we estimate
a shock radius $R_s$ $\sim$ 0.08--0.11~pc ($\sim$ 3\as--5\as \ at 5~kpc).
The pulsar wind parameter $\sigma$ is $\sim$~(4--11)~$\times$~10$^{-4}$,
indicating a particle dominated wind.
In Table~3, we summarize the inferred parameters of the
putative pulsar in \g21, in comparison with the Crab and 3C~58 pulsars.

\begin{table}[h]
\begin{center}
\caption{Inferred parameters of 
the putative pulsar in G21.5-0.9, in comparison with the
Crab and the newly discovered pulsar in 3C~58 (Slane \ea, this proceedings).}
\begin{tabular}{c|ccc|}

        &       \g21    &       Crab & 3C~58 \\ \tableline
Distance (kpc) &       5       &       2  & 2.6 \\
$L_X$ (0.5--10 keV) (10$^{35}$ erg~s$^{-1}$) &        3.3$D_5^2$      & 210 & 0.3 \\
$\dot{E}$ (10$^{37}$ erg~s$^{-1}$) &
3--6    &  47 & 2.6 \\
$\sigma$ &      (4--11)$\times$10$^{-4}$   &     3$\times$10$^{-3}$
& -- \\
$P$ (ms)\tablenotemark{a} &
$144 \tau_3^{-0.5} \dot{E}_{37}^{-0.5}$ & 33 & 67 \\
$B_0$ ($10^{13}$ Gauss) & $1.1~\tau_3^{-1}\dot{E}_{37}^{-0.5}$      & 0.4 & 0.36 \\
\end{tabular}
\tablenotetext{a}{
 $\tau_3$ is the age in units of
3 kyr; $\dot{E}_{37}$ the spin-down energy
loss in units of 10$^{37}$ erg~s$^{-1}$}
\end{center}
\end{table}
\section{Conclusions and future work}
The extended component is non-thermal--in agreement with \xmm\
observations (Warwick \ea\ 2001).
The softening of the spectral index away from
the core could be explained by synchrotron losses.
Unlike other plerions, \g21 is the only candidate
whose X-ray size is bigger than its radio size.
Deep radio observations (recently performed with the VLA) will allow us to
search for the radio counterpart of the
extended X-ray component and the missing SNR shell.
Since \g21 is bright and heavily absorbed,
a significant fraction of its flux
would be scattered by dust (R. Smith, private communication).
Future work should include modeling a dust scattering
X-ray halo.

\end{document}